\documentclass{emulateapj} \usepackage{psfig} \usepackage{apjfonts}
\pdfoutput=1
\usepackage{graphicx}

\newcommand\beq{\begin{equation}}
\newcommand\eeq{\end{equation}}
\newcommand\lsim{\mathrel{\rlap{\lower4pt\hbox{\hskip1pt$\sim$}}
        \raise1pt\hbox{$<$}}}
\newcommand\gsim{\mathrel{\rlap{\lower4pt\hbox{\hskip1pt$\sim$}}
        \raise1pt\hbox{$>$}}}

\begin{document}
\title{Ohmic Dissipation in the Atmospheres of Hot Jupiters}

\author{Rosalba Perna,\altaffilmark{1} Kristen Menou\altaffilmark{2,3}
and Emily Rauscher\altaffilmark{2,3}}
\altaffiltext{1}{JILA and
Department of Astrophysical and Planetary Sciences, University of
Colorado, Boulder, CO, 80309} \altaffiltext{2}{Department of
Astronomy, Columbia University, 550 West 120th Street, New York, NY
10027} \altaffiltext{3}{Kavli Institute for Theoretical Physics, UCSB,
Santa Barbara, CA 93106}

\begin{abstract}
Hot Jupiter atmospheres exhibit fast, weakly-ionized winds. The
interaction of these winds with the planetary magnetic field generates
drag on the winds and leads to ohmic dissipation of the induced
electric currents. We study the magnitude of ohmic dissipation in
representative, three-dimensional atmospheric circulation models of
the hot Jupiter HD~209458b.  We find that ohmic dissipation can reach
or exceed 1\% of the stellar insolation power in the deepest
atmospheric layers, in models with and without dragged winds. Such
power, dissipated in the deep atmosphere, appears sufficient to slow
down planetary contraction and explain the typically inflated radii of
hot Jupiters. This atmospheric scenario does not require a top
insulating layer or radial currents that penetrate deep in the
planetary interior. Circulation in the deepest atmospheric layers may
actually be driven by spatially non-uniform ohmic dissipation. A
consistent treatment of magnetic drag and ohmic dissipation is
required to further elucidate the consequences of magnetic effects for
the atmospheres and the contracting interiors of hot Jupiters.
\end{abstract}

\section{Introduction}

The discovery of the first transiting extrasolar planet HD~209458b
(Charbonneau et al. 2000) has opened a new chapter for the study of
planetary bodies. This planet belongs to the class of hot Jupiters,
which are close-in gaseous giant planets thought to be tidally locked
to their host star.  In the last decade, a wealth of observations has
allowed direct investigations of the atmospheric and bulk properties
of these planets, while much theoretical work has been aimed at the
interpretation of these data (see, e.g., Showman et al. 2010, Burrows
\& Orton 2010, Baraffe et al. 2010 and Seager \& Deming 2010 for
recent reviews).  A property of hot Jupiters which has constituted a
longstanding puzzle is their anomalously large radii, which have
been interpreted as requiring that extra heat be deposited
in {the convective regions, or alternatively in the deep atmospheres
(at pressure of tens of bars, e.g. Guillot \& Showman 2002)} of these planets.

In recent years, progress in modeling the unusual atmospheric
circulation regime of hot Jupiters, with permanent day and night
sides, has also been made (e.g. Cooper \& Showman 2005; Dobbs-Dixon \&
Lin 2008; Showman et al. 2009; Rauscher \& Menou 2010; see Showman et
al. 2010 for a review).  These purely hydrodynamical models have
generally ignored the possibility that magnetic effects acting on the
weakly-ionized winds could significantly influence the circulation
pattern.

Recently, Perna et al. (2010) evaluated the level of magnetic drag on
a representative hot Jupiter atmospheric flow, and argued that it is
likely to provide an effective frictional mechanism to limit the
asymptotic speed of winds in these atmospheres. Batygin \& Stevenson
(2010) studied the role of ohmic dissipation in hot Jupiters, using a
prescribed zonal wind profile and a one-dimensional atmospheric
structure model, to argue that the typical amount of heat deposited
deep in these planets' convective zones is sufficient to explain
their inflated radii. In this {\em Letter}, we build on our previous
work on magnetic drag (Perna et al. 2010; hereafter PMR10) to
investigate the magnitude of ohmic dissipation in the atmospheres of
hot Jupiters, and its consequences for the dynamics and the thermal
evolution of these planets. We use specific three-dimensional
atmospheric circulation models of the planet HD~209458b. We compare
the amount of Ohmic heating expected for typical magnetic field
strengths with the extra heat required in the deep atmosphere to slow
down contraction according to planetary evolutionary models (Guillot
\& Showman 2002). While our calculations are specific to the case of
HD~209458b, our overall results are expected to hold more generally
for hot Jupiters with similar gravity, irradiation strength, and
magnetic field.

\section{Atmospheric Currents and Ohmic Dissipation}

\subsection{Atmospheric models and ionization balance}

The fiducial atmospheric circulation model used here was computed by
Rauscher \& Menou (2010) for HD~209458b, under the assumption of no
significant {magnetic} drag {on the atmospheric flow}. The
model describes the atmospheric flow in a frame that is rotating with
the bulk planetary interior. The meridional and zonal wind speeds in
the atmosphere, as well as its thermodynamic variables, are returned
at each grid location in the three-dimensional model
atmosphere. Location is identified by the angular spherical
coordinates $(\theta,\phi)$ and pressure, $p$, for the vertical
coordinate, The model bottom is located at 220 bar while the top level
is set at a pressure of 1 mbar.  In addition to this drag-free model,
we also perform some of our calculations for the model with strongest
drag described in PMR10. Since, apart from wind drag, these two models
are identical, this allows us to evaluate the consequences for ohmic
dissipation of an atmospheric flow with significantly dragged winds.

In our circulation models, the local heating/cooling rate (energy per
unit mass) is modeled as Newtonian (linear) relaxation, $Q_T =
({T_{\rm eq}(p,\theta,\phi) - T})/{\tau_{\rm rad}(p)}$, where
$\tau_{\rm rad}$ represents the radiative timescale on which the local
temperature $T$ relaxes to the prescribed equilibrium profile $T_{\rm
eq}$.  {The nature of the atmospheric circulation obtained with
this simplified forcing compares well to what is obtained with more
realistic forcing (e.g., Showman et al. 2009).}  Following Cooper \&
Showman (2005), Rauscher \& Menou (2010) relied on the work of Iro et
al. (2005) to implement the detailed profiles of this radiative
forcing.  The atmosphere is divided into actively forced layers, above
the 10 bar level, and ``inert'' layers below.  The radiative forcing
is assumed to be negligible in the inert layers, which corresponds to
$\tau_{\rm rad}\rightarrow \infty$. The active layers, on the other
hand, are forced on a finite radiative timescale, $\tau_{\rm
rad}(p)$. The local atmospheric temperatures obtained after model
relaxation hover at about 1800 K in the deepest levels, while in the
upper levels they range from about 500 K on the night side to about
1500 K on the day side (see Rauscher \& Menou 2010 for details\footnote{
 While the simulations by Rauscher \& Menou (2010) adopt an
approximate treatment for radiative transfer, other simulations by
Showman et al. (2009) with explicit radiative transfer find rather
similar results for the day-night temperature gradient and wind
pattern under comparable physical conditions. However, it should be
recognized that all these models are subject to some uncertainties for
the development of the deep atmospheric winds due to the very long
integration times needed for spin-up.}).

At these temperatures, the primary source of free electrons stems from
thermally ionized alkali metals with low first ionization potentials:
Na, Al, K.  For simplicity here (and consistently with PMR10), we
adopt an approximation to Saha's equation (Balbus \& Hawley 2000)
which assumes potassium to be the dominant contributing species:
\begin{eqnarray}
x_e\equiv\frac{n_e}{n_n}&=&6.47\times 10^{-13}\left(\frac{a_K}{10^{-7}}\right)^{1/2}
\left(\frac{T}{10^3}\right)^{3/4}\nonumber \\ 
&\times&\left(\frac{2.4\times 10^{15}}{n_n} 
\right)^{1/2}\frac{\exp(-25188/T)}{1.15\times 10^{-11}}\;.
\label{eq:xe}
\end{eqnarray}  
Here $n_e$ and $n_n$ are respectively the electron and neutral number
densities (in cm$^{-3}$) and $a_K \simeq 10^{-7}$ is the potassium
solar abundance. As discussed in PMR10, equation~(\ref{eq:xe}) is a
good approximation as long the resulting ionization fraction, $x_e$,
is $\ll a_K$; this condition is satisfied in our atmosphere models.

We assume that the gas is overall neutral, which implies an equality
between the electron number density $n_e$ and the ionic one $n_i$.
The electrical conductivity and associated resistivity are given by
{(see also Laine et al. 2008)}
\begin{equation} \sigma_e =\frac{n_e e^2}{m_e n_n \left<\sigma {\rm v}\right>_e}\;\;\;{\rm
  and}\;\;\; \eta = \frac{c^2}{4\pi\sigma_e}\;,
\label{eq:sig-eta}
\end{equation}
respectively, with the collision rate between electrons and neutrals
approximated as (Draine et al 1983)
\begin{equation}
\left<\sigma {\rm v}\right>_{e}=10^{-15}
\left(\frac{128kT}{9\pi m_e}\right)^{1/2}{\rm cm}^3 {\rm s}^{-1}\;.
\label{eq:rate}
\end{equation}

\begin{figure}
\centering 
\includegraphics[scale=0.45]{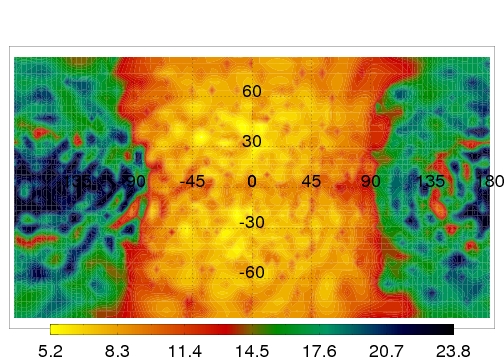}\\
\includegraphics[scale=0.45]{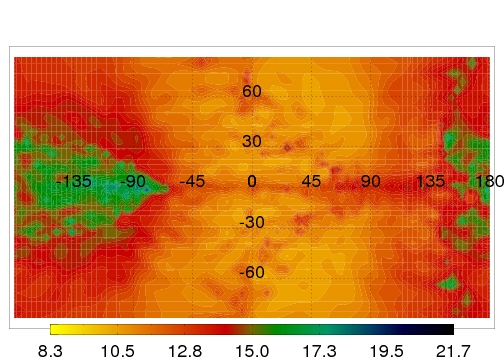}\\
\includegraphics[scale=0.45]{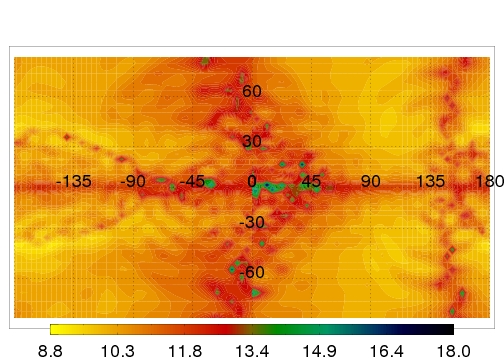}\\
\includegraphics[scale=0.45]{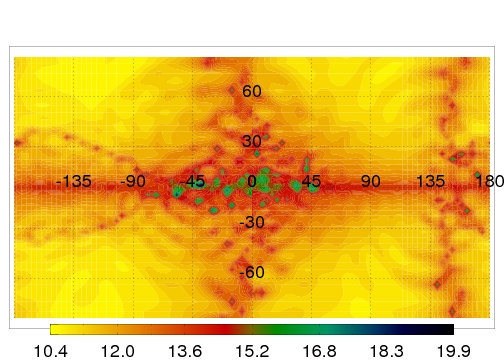}
\caption{Cylindrical maps of Joule heating times ($\log \tau_J$, in
  sec), at four pressure levels in our fiducial, drag-free atmospheric
  model.  The sub-stellar point is centered at longitude and latitude
  zero. From top to bottom, Joule heating times are shown at 1 mbar,
  50 mbar, 2 bar, 90 bar. Ohmic dissipation is not spatially uniform
  and it can dominate over heating by stellar insolation at deep
  enough pressure levels ($p \gsim 2$ bars), where a new form of
  Joule-driven circulation could emerge. The calculation was performed
  for $B=3$~G, but note that ohmic dissipation increases steeply with
  magnetic field strength ($\tau_J\propto B^{-2}$).}
\label{fig:times}
\end{figure}

\subsection{Induced currents and Ohmic dissipation} \label{ssec:cur}

The first step towards the computation of currents involves an
estimate of the importance of various non-ideal MHD terms in the full
induction equation for the weakly-ionized medium.  For the atmospheric
models under consideration here, PMR10 found that, for a surface
magnetic field strength of 3 G, the Hall term is completely negligible
throughout the flow, while the resistive term largely dominates over
the ambipolar diffusion term everywhere except possibly in the model
uppermost levels. Therefore, to a good approximation, the induction
equation can be considered as purely resistive. Assuming that zonal
winds are dominant, one needs only to consider the toroidal component
of the induction equation
\begin{eqnarray}
\frac{\partial B_\phi}{\partial t} &=& r\sin\theta\left[\frac{\partial\Omega}{\partial r}  
B_r + \frac{1}{r}\frac{\partial\Omega}{\partial \theta}  
B_\theta\right] + \frac{1}{r}\frac{\partial}{\partial r}\left[\eta\frac{\partial}{\partial r}
(rB_\phi)\right] \nonumber \\
&+& \frac{1}{r^2}\frac{\partial}{\partial\theta}\left[\frac{\eta}{\sin\theta}
\frac{\partial}{\partial\theta}(\sin\theta B_\phi)\right]\;,
\label{eq:indphi}
\end{eqnarray}
where $\Omega={\rm v}_\phi (r \sin\theta)^{-1}$ is the local angular
velocity of the flow, $r$ is the {radial spherical} coordinate, {and
the magnetic field is assumed to be an axisymmetric dipole}.

In PMR10, we also argued that, to leading order, the {latitudinal}
component of the current induced by the zonal flow should be
dominant. This assumption could be verified in more general versions of
our models.  {Since the magnetic Reynolds number is $R_m\ll 1$ for the
bulk of the flow, no significant dynamo action is expected in the
atmosphere. The dipolar planetary field $(B_r, B_\theta)$ is
maintained by currents in the interior of the planet, while the zonal
flows separately induce a $B_\phi$ component from the dipolar one in
the superficial atmospheric layers.}  Under these conditions (Liu et
al. 2008), the resulting steady-state {latitudinal} current can be
computed as:
\begin{eqnarray}
j_\theta(r,\theta,\phi)&=&-\frac{c \sin\theta}{4\pi r\eta(r,\theta,\phi)}
\int_r^R dr' {r'}^2\left( \frac{\partial \Omega}{\partial r'}B_r  
+\frac{1}{r'}\frac{\partial \Omega}{\partial\theta} B_\theta\right) \nonumber \\ 
&+& \frac{R\eta(R,\theta,\phi)}{r\eta(r,\theta,\phi)} j_\theta(R,\theta,\phi)\;,
\label{eq:j}
\end{eqnarray}
where the last term includes a boundary current in the uppermost
modeled level, $j_\theta(R,\theta,\phi)$.  Lacking information about
the nature of currents possibly flowing from regions above the modeled
atmospheric layers, we set this boundary current to zero for
simplicity. As discussed in PMR10, this unknown boundary current
represents an important source of uncertainty in our modeling, but,
unless near cancellations occur, additional boundary currents could in
principle contribute to even stronger ohmic dissipation than estimated
here.

The ohmic power per unit volume dissipated locally is readily computed
as (e.g., Liu et al. 2008)
\begin{equation}
Q_J(r,\theta,\phi) = \frac{[{j_\theta(r,\theta,\phi)}]^2}{\sigma_e(r,\theta,\phi)}\; .
\label{eq:QJ}
\end{equation}

\subsection{Joule-driven Circulation}

We first examine the possibility that spatially non-uniform ohmic
dissipation drives a circulation in regions of the atmosphere where it
is comparable or stronger than heating due to stellar insolation. To
permit a direct comparison between the two heating sources, we
find it convenient to define a typical local timescale associated with
Joule heating, which is deduced from the energy equation \beq
\rho\,C_p\,\frac{dT}{dt} = Q_J\,,
\label{eq:Cp}
\eeq
so that the Joule heating time is
\beq
\tau_J\sim \frac{\rho\, C_p\, T}{Q_J}\;. 
\label{eq:tauJ}
\eeq The specific heat of the gas in our model atmospheres is
$C_p=1.43\times 10^8$ erg g$^{-1}$ K$^{-1}$.

Fig.~1 shows cylindrical maps of Joule heating times for a nominal
surface dipolar field of 3 G, at four depths in the atmospheric flow
of our fiducial, drag-free model.  In the uppermost levels, the Joule
heating times span almost 20 orders of magnitude, with the shortest
times found on the day side. This wide span reflects the large
variations in resistivity, $\eta$, between the day and the night
sides. On the day side, which has higher temperatures, $\eta$ is
considerably smaller than on the night side, yielding larger currents,
which in turn result in larger $Q_J$ and correspondingly smaller
$\tau_J$ values. On the other hand, deep in the inert layers, which
are little affected by stellar irradiation, Joule heating times span a
more modest range of values, $\sim 10^9-10^{12}$~s over a large
fraction of the flow. Longer timescales along the equator in the
  deeper levels are mostly the result of geometric effects: in
  Eq.(\ref{eq:j}), the term $(\partial\Omega/\partial r)B_r$ generally
  dominates over $1/r (\partial\Omega/\partial\theta) B_\theta$. Since
  $\Omega\propto\sin^{-1}\theta$ and $B_r\propto \cos\theta$, the net
  result is $j\propto\cos\theta$, and hence $\tau_J\propto
  \cos^{-2}\theta$.  For a non-axisymmetric dipolar field, the
  anisotropy pattern would be different.

To evaluate the possibility of Joule-driven circulation, we
compare local Joule heating times with a representative heating time
associated with stellar irradiation, $\tau_{\rm irr}$. For consistency
with the Newtonian forcing scheme used in our atmospheric models, we
adopt a simple downward insolation flux that approximately matches the
absorption properties of the one-dimensional models computed by Iro et
al. (2005) at a few bars level. The flux is taken to obey $F(p)=F_0
\exp[-4.5 p/(2 {\rm bar})]$, where $F_0=2.23\times 10^8$ { erg}{
cm}$^{-2}$ ${ s}^{-1}$ is the incident flux at the model top for HD
209458b, after geometric dilution by a factor $1/4$ as is customary
for 1D-averaged models (e.g., Hansen 2008).  A stellar irradiation
heating timescale, $\tau_{\rm irr}$, is then computed following
Eq.~(\ref{eq:tauJ}), with a heating rate taken as the vertical
divergence of the stellar irradiation flux, $Q_{\rm
irr}=-dF(r)/dr$. This allows us to evaluate $\tau_{\rm irr}$ simply as
a function of the local pressure, pressure scale height and
density. For example, at $p=2$ bar, using $\rho \sim 5.5 \times
10^{-5}$~g~cm$^{-3}$, we estimate $\tau_{\rm irr} \sim 5 \times 10^7
$~s. At $p=10$ bar, where the stellar insolation flux has been very
strongly attenuated already (Iro et al. 2005), we estimate $\tau_{\rm
irr} \sim 6 \times 10^{15}$~s for $\rho \sim 3 \times
10^{-4}$~g~cm$^{-3}$. 
Deeper in the atmosphere, the insolation flux is
further attenuated (exponentially so) and heating by insolation
becomes largely inconsequential.

A comparison between values for $\tau_{\rm irr}$ and the detailed maps
of $\tau_J$ values in Fig.~1 shows that, for the nominal $B=3$~G used
in our $\tau_J$ calculation, irradiation dominates over Joule heating
($\tau_{\rm irr} \ll \tau_J$) everywhere {on the day side} of the
atmosphere at pressure levels above a few bars.  {In these upper regions, we
expect the atmospheric thermal structure to be largely unaffected by
the extra Ohmic heating.}  Deeper than a few bars, however, Joule
heating times start becoming comparable to or shorter than the typical
heating time associated with stellar insolation, over significant
regions of the atmospheric flow. Eventually, at levels deeper than
about 10 bar, ohmic dissipation easily dominates over stellar
insolation over much of the atmosphere.  Since this extra source of
heating in the deeper regions of the atmosphere is spatially
non-uniform (Fig.~1), it should lead to some form of Joule-driven
circulation in the ``inert'' layers located well below the
radiatively-forced ``weather'' layers. This conclusion is largely
independent of the specific stellar insolation model adopted here
since the existence of radiatively inactive layers at levels deeper
than a few bars is a rather generic property of irradiated hot Jupiter
atmospheres (e.g., Hansen 2008). However, it remains to be seen what
the nature of such Joule-driven circulation might be given the uneven
heating pattern shown in Fig.~1, the non-uniformity of stellar
insolation from the day-side to the night-side and the presence of an
additional, spatially uniform net flux emerging from the deep
planetary interior.  Finally, it is worth remembering that, since
$\tau_J\propto B^{-2}$, a stronger field could produce a substantially
more dominant ohmic dissipation than estimated above for
$B=3$~G\footnote{The magnetic field strengths of hot Jupiters are 
unconstrained from an observational point of view. 
In the case of Jupiter, the measured field  is 14~G
at the pole and 4.2~G around the equator. Arguments (e.g. Christensen et al. 2009)
suggest that the field strength should scale with the square root of the
density and the spin frequency, which could make the $B$ field in hot Jupiters
somewhat weaker. Calculations of the internal structure and convective motions
of giant planets (Sanchez-Lavega 2004) yield surface magnetic fields $\sim 1-2$~G
for synchronized hot Jupiters.}.

\subsection{Inflating Hot Jupiters}

Another potentially important consequence of ohmic dissipation is the
possibility that it contributes to the tendency for hot Jupiters to
have inflated radii (Batygin \& Stevenson 2010). To evaluate the
magnitude of this effect on the basis of our three-dimensional
atmospheric models, we compute the cumulative ohmic power dissipated
in successive atmospheric shells, from the top level at a pressure
$p_t$ of 1 mbar, down to a pressure $p(r)$ within the atmosphere; this
is readily obtained by integration of Eq.(\ref{eq:QJ}),
\begin{equation}
P_J(p) = \int_{r(p)}^{r(p_t)}dr' {r'}^2 \int_0^{2\pi} d\phi\int_0^{\pi} d\theta\;
\frac{[{j_\theta(r',\theta,\phi)}]^2}{\sigma_e(r',\theta,\phi)}\;.
\label{eq:PJ}
\end{equation}
{Note that the power as computed above includes a horizontal average
over the day and night sides. This averaging washes out the horizontal
features in Joule heating shown in Fig.~1, an issue deserving further
attention in future multi-dimensional models.}  Fig.~2 displays this
ohmic power as a function of pressure for our fiducial, drag-free
atmospheric model and for the model with strongest drag described in
PMR10, with zonal wind speeds typically reduced by about
$30-50\%$. For each model, we display ohmic power for two different
values of the magnetic field strength adopted in the ohmic dissipation
calculation, $B=3$~G, and $B=10$~G. This allows us to separately
explore the effects of drag on the zonal winds and of the magnetic
field strength on the resulting magnitude of ohmic dissipation. This
is a useful exercise until models with a self-consistent treatment of
magnetic drag and ohmic dissipation become available.

Fig.~2 shows that much of the dissipated ohmic power builds up at
pressure levels from a few bars to several tens of bars. The power
scales as $B^2$ and, relative to the drag-free model, it is reduced by
a factor $\sim 3$ in the circulation model with strongly dragged
winds, at fixed magnetic field strength. Since the stellar insolation
power received by HD 209458b is $\sim 3 \times 10^{22}$~W, Fig.~2
shows that the total ohmic power dissipated in these model atmospheres
typically approaches or exceeds 1\% of the stellar insolation power.
Notice that, while the Ohmic power in Fig.2 is integrated down to
the lowest grid points ($\sim 200$~bar) available in the model of
Rauscher \& Menou (2010), it is not yet seen to saturate at those
levels. This is because, while the zonal velocities decrease with
depth, they are still non-zero in the deepest, 'inert' model layers.
It is in fact not a priori clear where exactly the transition to the
convection zone should be located in hot Jupiters and this
constitutes an uncertainty for Joule heating models of the deepest
atmospheric layers.

Overall, our results on the dissipated ohmic power are consistent with
the estimates made by Batygin \& Stevenson (2010) on the basis of a
more idealized, one-dimensional atmosphere model, using a
parametrized zonal wind profile. Like them, we find that a large
fraction of the ohmic power is dissipated in the deeper atmospheric
levels, below $\sim 10$ bar.
\begin{figure}
\centering
\includegraphics[scale=0.45]{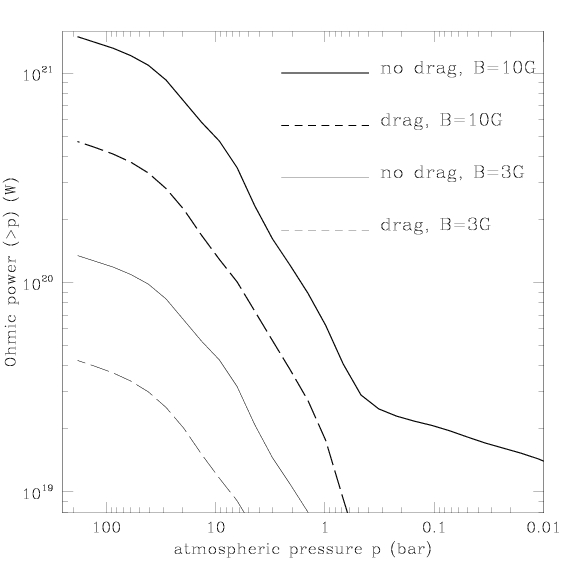}
\caption{Cumulative ohmic power dissipated above pressure level $p$,
  in four different models for HD 209458b. The solid lines correspond
  to the drag-free model described in Rauscher \& Menou (2010) while
  the dashed lines correspond to the model with strongest drag, and
  reduced winds, described in Perna et al. (2010). Curves are labeled
  with the value of the magnetic field adopted for the ohmic
  dissipation calculation. The ohmic power can reach or exceed 1\% of
  the stellar insolation power, $\sim 3 \times 10^{22}$~W, in the
  deepest atmospheric layers, in models with and without dragged
  winds. }
\label{fig:power}
\end{figure}
Contrary to Batygin \& Stevenson (2010), however, we do not solve for
radial currents in our models, only meridional ones (Eq.~\ref{eq:j}),
and our results are thus largely independent of the presence or
absence of an insulating layer high in the atmosphere. In fact, we
find that an insulating layer is present on the nightside in our
three-dimensional models, but not necessarily on the much hotter, less
resistive dayside. This raises the possibility that, in more detailed
three-dimensional MHD models, current loops could actually close high
in the atmosphere, rather than deep in the planetary interior as
conjectured by Batygin \& Stevenson (2010). While our results, relying
on leading-order meridional currents in the atmospheric region alone,
are presumably not strongly sensitive to these conditions affecting
radial currents, it remains to be seen how currents would flow in more
realistic non-axisymmetric and three-dimensional models.

Various authors have evaluated the extra power that needs to be
continuously dissipated deep in the convective interior of hot
Jupiters to explain their inflated radii (e.g., Gu, Bodenheimer \& Lin
2004, Burrows et al. 2007, Ibgui et al. 2009). While Batygin \&
Stevenson (2010) emphasize such a deep deposition scenario, our own
ohmic dissipation models say little about this scenario since we only
calculate currents and ohmic dissipation in the superficial
atmospheric region of the planet.  However, another means by which hot
Jupiter radii can be inflated is by slowing down their rate of
contraction, through modifications to the thermal structure of their
overlaying atmospheres which act as boundary conditions for the
cooling isentropic interiors. Indeed, Guillot \& Showman (2002) argue
that, if $\sim 1\%$ of the stellar insolation flux were deposited at
pressures of tens of bars deep in the atmosphere, this would slow down
cooling sufficiently to explain the inflated radii of hot
Jupiters. While Guillot \& Showman (2002) suggested that a downward
flux of kinetic energy could in principle achieve such energy
deposition, our Fig.~2 indicates that this is in fact naturally
achieved by ohmic dissipation in our atmospheric models, with or
without drag, for a magnetic field strength $B \gsim 10$~G.

More specifically, Guillot \& Showman (2002) describe an evolutionary
model in which $2.4 \times 10^{20}$~W are deposited at a location
centered around 21 bar in the atmosphere of HD 209458b, which can
explain its observed inflated radius. This is achieved by the two
ohmic dissipation models represented by the upper dashed and solid
lines in our Fig.~2. In fact, since even more ohmic power is
dissipated deeper in, we anticipate that models with fields even
weaker than $10$~G, possibly as low as 3~G, will be able to meet the
inflated radius requirement for HD 209458b. This leads us to conclude
that ohmic dissipation deep in the atmospheres of hot Jupiters, which
indirectly taps into the kinetic energy of dragged winds driven by
stellar insolation higher up, is a promising scenario to explain the
inflated radii of hot Jupiters.

\section{Summary and Conclusions}

In this {\em Letter}, we have computed the rate of ohmic dissipation
in representative, three-dimensional atmospheric circulation models of
the hot Jupiter HD~209458b. We find that, for a fiducial magnetic
field strength of 3~G, ohmic dissipation starts dominating over
stellar insolation heating at levels deeper than a few bars. The
spatial non-uniformity of this extra source of heat could induce
Joule-driven circulation in the deep layers traditionally considered
as ``inert''. For a magnetic field strength $\gsim 10$~G, our models
also indicate that enough heat is deposited at pressures of several
tens of bars to slow down cooling sufficiently that the inflated radii
of hot Jupiters can be explained.

Our results hence suggest that magnetic interactions in hot Jupiter
atmospheres play a fundamental coupling role for the dynamics and the
thermal evolution of these planets. As such, our work calls for the
problem to be treated more consistently: magnetic drag affects wind
speeds and induces currents, while the ohmic dissipation of these
currents can generate a deep atmospheric circulation which could, in
turn, feedback on the circulation higher up.  Indeed, the extra
heat source might also enhance convection in the night side, which
might in turn enhance cooling.  These various ingredients will have
to be incorporated consistently in circulation models for a better
assessment of their influence on the structure and the evolution of
hot Jupiters. Some diversity may naturally arise from variations in
the magnetic field strength and geometry of different planets.

\acknowledgements 

We thank Tristan Guillot for useful discussions, and Adam Burrows, Jeremy
Goodman and the referee Douglas Lin for helpful comments on our manuscript. 
This work was supported in part by the National Science Foundation under Grant
No. PHY05-51164.

\end{document}